\documentclass[showpacs,twocolumn,prd,nofootinbib]{revtex4-1}
 \usepackage{color,graphicx,epsfig}
 \usepackage{amsmath,amssymb,dsfont,epsfig,graphicx,xcolor}
 \usepackage{ifpdf}
 \usepackage{amsmath}
 \usepackage{bm}
 \usepackage{color}
 \usepackage[english]{babel}
 \usepackage{graphicx}%
 \usepackage{amsfonts}%
 \usepackage{amssymb}
 \usepackage{braket}
 \usepackage{hyperref}
 \usepackage{epstopdf}
 \graphicspath{{Plots/}}

\definecolor{nicered}{rgb}{0.7,0.1,0.1}
\definecolor{nicegreen}{rgb}{0.1,0.5,0.1}
\definecolor{violet}{rgb}{0.7,0.3,0.3}
\hypersetup{colorlinks,citecolor= nicegreen,linkcolor= nicered}

\newcommand{\nc}{\newcommand}
\nc{\non}{\nonumber}
\nc{\hc}{\hbox {H.c.}}
\nc{\noi}{\noindent}
\nc{\barx}{\bar{x}}
\nc{\pbarn}{\;\hbox {pb}}
\nc{\fbarn}{\;\hbox {fb}}

\nc{\hsp}{\hspace{0.5cm}}
\nc{\lsp}{\hspace{1cm}}
\nc{\Lsp}{\hspace{2cm}}
\nc{\LLsp}{\lsp\lsp}
\nc{\lra}{\longrightarrow}
\nc{\p}{\prime}
\nc{\sgn}{\text{sgn}}
\nc{\ph}{\varphi}
\nc{\op}{{\cal O}}
\nc{\eq}{\text{Eq.~}}

\nc{\beq}{\begin{equation}}  \nc{\eeq}{\end{equation}}
\nc{\bea}{\begin{eqnarray}}  \nc{\eea}{\end{eqnarray}}
\nc{\baa}{\begin{array}}     \nc{\eaa}{\end{array}}
\nc{\bit}{\begin{itemize}}   \nc{\eit}{\end{itemize}}
\nc{\ben}{\begin{enumerate}} \nc{\een}{\end{enumerate}}
\nc{\bce}{\begin{center}}    \nc{\ece}{\end{center}}
\nc{\bpm}{\begin{pmatrix}}   \nc{\epm}{\end{pmatrix}}
\nc{\bvt}{\begin{verbatim}}  \nc{\evt}{\end{verbatim}}

\def\mysection#1{{{\bf #1}.~}}
\begin{document}

\def\LjubljanaFMF{Faculty of Mathematics and Physics, University of Ljubljana,
 Jadranska 19, 1000 Ljubljana, Slovenia }
\def\LjubljanaIJS{Jo\v zef Stefan Institute, Jamova 39, 1000 Ljubljana, Slovenia}
\def\Orsay{Universit\'e Paris-Saclay, CNRS/IN2P3, IJCLab, 91405 Orsay, France}

\title{Implications of $b\to s\mu\mu$ Anomalies \\for Future Measurements of $B \to K^{(*)} \nu \bar \nu$ and $K\to \pi \nu \bar \nu$}

\author{S\'ebastien Descotes-Genon}
\email[Electronic address:]{sebastien.descotes-genon@ijclab.in2p3.fr} 
\affiliation{\Orsay}

\author{Svjetlana Fajfer}
\email[Electronic address:]{svjetlana.fajfer@ijs.si} 
\affiliation{\LjubljanaIJS}
\affiliation{\LjubljanaFMF}
 
\author{Jernej~F.~Kamenik}
\email[Electronic address:]{jernej.kamenik@cern.ch} 
\affiliation{\LjubljanaIJS}
\affiliation{\LjubljanaFMF}

\author{Mart\'in Novoa-Brunet}
\email[Electronic address:]{martin.novoa@ijclab.in2p3.fr} 
\affiliation{\Orsay}
\begin{abstract}
We investigate the consequences of deviations from the Standard Model observed in $b\to s\mu\mu$ transitions for flavour-changing neutral-current processes involving down-type quarks and neutrinos. We derive the relevant Wilson coefficients within an effective field theory approach respecting the SM gauge symmetry, including right-handed currents, a flavour structure based on approximate $U(2)$ symmetry, and assuming only SM-like light neutrinos.  We discuss correlations among  $B \to K^{(*)} \nu \bar \nu$  and $K\to \pi \nu \bar \nu$ branching ratios
in the case of linear Minimal Flavour Violation and in a more general framework, highlighting in each case the role played by various New Physics scenarios proposed to explain $b\to s\mu\mu$ deviations.
\end{abstract}

\maketitle

\mysection{Introduction}
Recent experimental data in $B$ physics hint toward deviations from Lepton Flavour Universality (LFU) in semi-leptonic decays~\cite{Bifani:2018zmi}
\beq
R_K= \frac{\mathcal B(B\to K \mu\mu)}{\mathcal B(B\to K ee)}     \,, \quad R_{K^*} =   \frac{\mathcal B(B \to K^* \mu\mu)}{\mathcal B(B \to K^*  ee)} \,,
\eeq
as measured by LHCb~\cite{Aaij:2014ora,Aaij:2019wad,Aaij:2017vbb} at significances from 2.3$\sigma$ to 2.6$\sigma$. Belle has also recently reported measurements of $R_K$~\cite{Abdesselam:2019lab} and $R_{K^*}$~\cite{Abdesselam:2019wac} in agreement with LHCb measurements, but with much larger uncertainties.
In addition to these LFU ratios, LHCb data exhibit deviations close to $3\sigma$ from the Standard Model (SM) expectation in the $P^\prime_5$ angular observable of $B \to K^*\mu\mu$ decay~\cite{Descotes-Genon:2013vna}, and milder deviations are also seen in branching ratios of $b\to s\mu\mu$ exclusive decays~\cite{Aaij:2013qta,Aaij:2014pli,Aaij:2015oid,Aaij:2020nrf,Aaij:2013aln}. Deviations are also hinted at in Belle data for $B\to K^*\mu\mu$~\cite{Abdesselam:2016llu,Wehle:2016yoi}.

These deviations can be interpreted model-independently in terms of specific contributions to the effective weak hamiltonian (see e.g. Ref.~\cite{Buchalla:1995vs, Buras:1998raa}) at the scale $m_b$
\beq
\mathcal H_{\rm eff} \ni - \frac{4 G_F}{\sqrt 2} V_{tb} V^*_{ts} \sum_{i,\ell}^{} \left[ C^\ell_i \mathcal O^\ell_i + C^\ell_{i^\prime} \mathcal O^\ell_{i^\prime} \right]\,,
\label{eq:Heff}
\eeq
where the relevant long-distance operators are
\beq 
\mathcal O^{\ell}_9 = \frac{\alpha_{}}{4\pi} \left( \bar s \gamma_\mu P_L b \right) \left( \bar \ell \gamma^\mu \ell \right), ~ \mathcal O^{\ell}_{10} = \frac{\alpha_{}}{4\pi} \left( \bar s \gamma_\mu P_L b \right) \left( \bar \ell \gamma^\mu \gamma_5 \ell \right)\,,
\eeq
while chirality-flipped operators $\mathcal O^\ell_{9^\prime,10^\prime}$ are obtained from the above expressions with the replacement $P_L \to P_R$  with $P_{L,R} =(1\mp\gamma_5)/2 $. The global fits to $b\to s\ell\ell$ data (see Ref.~\cite{Alguero:2019ptt} and references therein) show that these deviations exhibit a consistent pattern favouring a significant additional New Physics (NP) contribution to the short-distance Wilson coefficient $C^\mu_9$ (of the order of 25\% of the SM contribution) together with smaller contributions to $C^\mu_{10}$ and/or $C^\mu_{9'}$. Among the scenarios improving by 5$\sigma$ or more the description of the data compared to the SM, one can find the one-dimensional scenarios with the best-fit point and the 68\% confidence intervals, favoured on the basis of their pulls with respect tot the Standard Model~\cite{Alguero:2019ptt}:
\begin{eqnarray}
C^{\mu,{\rm NP}}_9 &:& -1.03 \quad [-1.19,-0.88] \quad 6.3\sigma\,,\\
C^{\mu,{\rm NP}}_9=-C^{\mu,{\rm NP}}_{10} &:& -0.50 \quad [-0.59,-0.41] \quad 5.8\sigma\,,\\
C^{\mu,{\rm NP}}_9=-C^{\mu,{\rm NP}}_{9'} &:& -1.02 \quad [-1.17,-0.87] \quad 6.2\sigma\,.
\end{eqnarray}
Two-dimensional scenarios achieving similarly high pulls with respect to the SM are obtained for NP contributions to $(C^\mu_{9},C^\mu_{10})$ and $(C^\mu_{9},C^\mu_{9^\prime})$. Smaller contributions to electron operators are allowed by the data, but not required to achieve a good description. Similar results have been obtained by other groups performing such fits choosing  different theoretical inputs and experimental subsets and different statistical frameworks~\cite{Aebischer:2019mlg,Ciuchini:2019usw,Arbey:2019duh}. 

The SM neutrinos reside in the same leptonic weak doublets as the left-handed charged leptons. Therefore, decay modes with neutrinos in the final state offer complementary probes of NP.  In particular, decays $B\to h_s \nu \bar \nu$, with $h_s$  standing  for  hadronic states of unit strangeness, are known for their NP sensitivity~\cite{Altmannshofer:2009ma}.  In the SM,  branching ratios are found to be $\mathcal B(B\to K^{(*)}  \nu \bar \nu)_\mathrm{SM}=(9.6 \pm 0.9)\times 10^{-6}$ and $\mathcal B(B^+ \to K^+  \nu \bar \nu)_\mathrm{SM}= (5.6\pm 0.5)\times 10^{-6}$ \cite{Kou:2018nap}. The Belle collaboration has produced limits at 90\% Confidence Level (CL): $\mathcal B(B^0\to K^{*0}\nu \bar \nu)_\mathrm{exp}< 1.8 \times 10^{-5}$,   $\mathcal B(B^+\to K^{*+}\nu \bar \nu)_\mathrm{exp}< 6.1 \times 10^{-5}$ and  $\mathcal B(B^+ \to K^+  \nu \bar \nu)_\mathrm{exp}< 1.9\times 10^{-5}$ \cite{Grygier:2017tzo}. The Belle II collaboration plans to  observe these three decay modes  with about $10$ ab$^{-1}$ of data, while the sensitivities to the SM branching ratio will reach a precision of about $10\%$ with $50$ ab$^{-1}$~\cite{Kou:2018nap}. 

In the kaon sector $K \to \pi \nu \bar \nu$  decays arguably offer the best sensitivity to NP~\cite{Buras:1998raa}. The experimental limits at 90\% CL for the  branching ratios were obtained by the E949 experiment 
$ \mathcal B (K^+ \to \pi^+ \nu \bar\nu)_\mathrm{exp} < 3.35  \times 10^{-10}$ ~\cite{Artamonov:2008qb,Artamonov:2009sz} with a recent update  from the NA62 collaboration
$ \mathcal B (K^+ \to \pi^+ \nu \bar\nu)_\mathrm{exp} <  2.24 \times 10^{-10}$~\cite{Ruggiero},
while KOTO presented preliminary results indicating $\mathcal B (K_L \to \pi^0 \nu \bar\nu)_\mathrm{exp} = 2.1^{+2.0(+4.1)}_{-1.1(-1.7)} \times 10^{-9}$~\cite{Shinohara}.
The corresponding SM values are $\mathcal B(K^+ \to \pi^+ \nu \bar\nu)_\mathrm{SM} = (9.31 \pm0.76)\times 10^{-11} $ and 
$\mathcal B (K_L \to \pi^0 \nu \bar\nu)_\mathrm{SM} = (3.74\pm0.72) \times 10^{-11}$ ~\cite{Buras:2015qea}.  

We recall that these two branching ratios should obey the Grossman-Nir bound $\mathcal B (K_L \to \pi^0 \nu \bar\nu) \leq 4.3 \, \mathcal B (K^+ \to \pi^+ \nu \bar\nu) $, in which the numerical factor  results from the difference in the total decay widths  of $K_L$  and $K^+$, isospin breaking effects, and QED radiative corrections ~\cite{ Grossman:1997sk}.  In particular, the Grossman-Nir bound constitutes an additional very strong theoretical constraint on $\mathcal B (K_L \to \pi^0 \nu \bar\nu)$.
The experiment NA62 is planning to eventually measure the rate of $K^+ \to \pi^+ \nu \bar\nu$ with $O(10\%)$ precision~\cite{Ruggiero:2017hjh}.  For the neutral decay mode $K_L \to \pi^0 \nu \bar\nu$ KOTO and KLEVER also aim at making significant progress~\cite{Strategy:2019vxc} and resolving the current somewhat ambiguous situation with respect to possible NP effects in these modes~\cite{Kitahara:2019lws}. 

Precise results for all these rare semileptonic $b\to s$ and $s\to d$ transitions will allow to get much better insight into possible NP effects observed in $R_{K^{(*)}}$.  Particularly interesting is  the question whether NP is  only present in $b \to s$ transitions or also in other Flavour-Changing Neutral Currents (FCNC).  The  measurements of $s\to d \nu \bar \nu$  and $b\to s \nu \bar \nu$  rates will help to differentiate  among NP  models with different flavour and chiral structures in both the quark and lepton sectors.  
This issue has been already raised in many studies which mostly relied on particular models of NP \cite{Bordone:2017lsy,Kamenik:2017tnu,Fajfer:2018bfj,Altmannshofer:2009ma,Aebischer:2019blw}. The main goal of our  approach is to determine the impact of  $R_{K^{(*)}}$ on future measurements of $B \to K^{(*)} \nu \bar \nu$ and $K\to \pi \nu \bar \nu$ in a general effective theory framework and to illustrate the potential correlations among these measurements.

\mysection{NP in semileptonic FCNC decays}
Possible heavy NP contributions should be written in terms of $SU(2)_L$ gauge invariant operators~\cite{DAmbrosio:2002vsn,Hurth:2008jc,Buttazzo:2017ixm}, e.g.
\begin{align}
& \mathcal L_{\rm eff.}   = \mathcal L_{\rm SM}  - \frac{1}{v^2} \lambda^q_{ij} \lambda^\ell_{\alpha\beta} \left[ C_T \left( \bar Q_L^i \gamma_\mu \sigma^a Q_L^i \right) \left( \bar L_L^\alpha \gamma^\mu \sigma^a L_L^\beta \right)  \right. \nonumber \\
  &   +  C_S \left( \bar Q_L^i \gamma_\mu  Q_L^i \right) \left( \bar L_L^\alpha \gamma^\mu  L_L^\beta \right)  + C'_{RL} \left( \bar d_R^i \gamma_\mu d_R^i \right) \left( \bar L_L^\alpha \gamma^\mu  L_L^\beta \right) \nonumber\\
  &  \left.  +  C'_{LR} \left( \bar Q_L^i \gamma_\mu  Q_L^i \right) \left( \bar \ell_R^\alpha \gamma^\mu  \ell_R^\beta \right)  + C'_{RR} \left( \bar d_R^i \gamma_\mu d_R^i \right) \left( \bar \ell_R^\alpha \gamma^\mu  \ell_R^\beta \right)
  \right]\,, \label{eq:ops}
\end{align}
where we choose to write the operators in the down-quark and charged lepton mass basis
$Q_L^i = (V^{\rm CKM *}_{ji}  u_L^j, d_L^i)^T$
and $L_L^\alpha = (U^{\rm PMNS}_{\alpha\beta}\nu_L^\beta,\ell^{\alpha}_L)^T$.  
Following Refs.~\cite{Buttazzo:2017ixm,Bordone:2017lsy,Cornella:2019hct} 
we assume that the same flavour structure encoded in $\lambda^q_{ij}$ and $\lambda^\ell_{\alpha\beta}$ holds for all operators. As it will become clear from the discussion below, this assumption actually does not result in any loss of generality of our main results. 
It also turns out to be beneficial to classify the NP flavour structure in terms of an approximate  $U(2)_{q=Q,D}$ flavour symmetry acting on quark fields, under which two generations of quarks form doublets, while the third generation is invariant. One can write 
${\bf q}\equiv (q_L^1, q_L^2) \sim ({\bf 2},{\bf 1})$, ${\bf d} \equiv (d_R^1,d_R^2) \sim ({\bf 1},{\bf 2})$ while $ d_R^3,q_L^3 \sim ({\bf 1},{\bf 1})$\,.
In the exact $U(2)_{q}$ limit only $\lambda^q_{33}$ and $\lambda^{q}_{11}=\lambda^q_{22} $ in Eq.~\eqref{eq:ops} are non-vanishing.
Since a specific pattern of $U(2)_{q}$ breaking (by the SM Yukawas) is required to accommodate the first two generation quark masses and the CKM matrix, there is an ambiguity in the definition of the singlet field with respect to the down-quark mass basis, which, if chosen arbitrarily, may still result in unacceptably large mixing among generations. To avoid excessive effects in neutral kaon oscillation observables,
we thus furthermore impose the {\it leading} NP $U(2)_q$ breaking to be aligned with the SM Yukawas, yielding a General Minimal Flavour Violating (GMFV)~\cite{Kagan:2009bn} structure
\begin{eqnarray}
\label{b3q}
 q_{L}^3 &= & \begin{pmatrix}V_{j b}^* u_L^j\\ b_L\end{pmatrix}+ \theta_q e^{i\phi_q} \left[ V_{td} \begin{pmatrix} V_{jd}^*  u_L^j\\ d_L  \end{pmatrix} +  V_{ts} \begin{pmatrix}V_{js}^*  u_L^j\\s_L  \end{pmatrix}\right].\nonumber\\
 \end{eqnarray}
 where $\theta_q$ and $\phi_q$ are fixed but otherwise arbitrary numbers. Therefore respecting the $U(2)_q$ symmetry with GMFV breaking one has
$d _{L}^3 =  b_L+ \theta_q e^{i\phi_q} \left( V_{td} d_L   + V_{ts}  s_L \right)$. The linear MFV limit ~\cite{Kagan:2009bn} is recovered by taking $\theta_q = 1$ and $\phi_q=0$ (taking $V_{tb}=1$). In (G)MFV,  right-handed FCNCs among down-type quarks are suppressed so that we may set $C'_{RL}=C'_{RR}=0$ then.
Departures from the (G)MFV limit may manifest through additional
 explicit $U(2)_q$ breaking effects appearing as $\lambda^q_{i\neq j} \neq 0$ and we normalise such effects by the $U(2)_q$ symmetric ($\lambda^q_{33}$) contribution by defining $r_{ij}={\lambda_{ij}^q}/{\lambda_{33}^q}$.

For the lepton sector we assume an approximate $U(1)^3_\ell$ symmetry (broken only by the neutrino masses) yielding $\lambda_{i\neq j}^\ell \simeq 0$\, as required by stringent limits on lepton flavour violation.  We consider here only (SM-like) left-handed neutrinos. As discussed in the Introduction, current (LFU) NP hints in rare semileptonic B decays only indicate significant non-standard effects in muonic final states. While a smaller effect in electrons is not excluded, $b\to s \tau^+ \tau^-$ transitions are at present only poorly constrained and could in principle exhibit even much larger deviations than those observed in $R_{K^{(*)}}$~\cite{Capdevila:2017iqn}.  However, the corresponding neutrino flavours are not tagged in  current and upcoming rare meson decay experiments. In order to correlate FCNC processes involving charged leptons and neutrinos we need to assume specific ratios of $U(1)_\ell^3$ charges ($\lambda^{\ell}$). In the following we will consider three well known examples from the existing literature:
\begin{enumerate}
\item The simplest $\lambda^{\ell}_{ee} =\lambda^{\ell}_{\tau\tau} = 0$ scenario implies significant NP effects only in muonic final states. Correspondingly only a single neutrino flavour $\nu$ (e.g. in the sums of Eqs.~\eqref{eq:BKnunu} and~\eqref{eq:Kpinunu}) receives NP effects. This is usually assumed in model-independent EFT analyses.

\item The anomaly-free assignment $\lambda^{\ell}_{\mu\mu} = -\lambda^{\ell}_{\tau\tau}$ and $\lambda^{\ell}_{ee} =0$ allows for gauging of the leptonic flavour symmetry and is thus well suited for UV model building~\cite{Altmannshofer:2014cfa, Crivellin:2016ejn}. In this case two of the neutrino flavours in Eqs.~\eqref{eq:BKnunu} and~\eqref{eq:Kpinunu} receive NP effects, equal in magnitude, but opposite in sign.

\item The hierarchical charge scenario $\lambda^{\ell}_{ee} \ll \lambda^{\ell}_{\mu\mu} \ll \lambda^{\ell}_{\tau\tau}$ is motivated by models of partial lepton compositeness and flavour models accounting for hierarchical charged lepton masses~\cite{Redi:2011zi, Niehoff:2015bfa}. In this case NP effects in Eqs.~\eqref{eq:BKnunu} and~\eqref{eq:Kpinunu} are again dominated by a single ($\tau$) neutrino flavour, however the effects can be much larger than indicated by the deviations in $R_{K^{(*)}}$. For concreteness in the following we consider $\lambda^{\ell}_{\tau\tau} / \lambda^{\ell}_{\mu\mu} = m_\tau / m_\mu$  and again neglect the small effects in $\lambda^{\ell}_{ee}$ for this scenario.
\end{enumerate}

The Wilson coefficients appearing in Eq.~\eqref{eq:Heff} can now be expressed compactly as $C_i^{\mu} = C^{\ell,{\rm SM}}_{i} + C^{\mu,{\rm NP}}_{i} $ where  to linear order in $U(2)_q$ breaking
\begin{eqnarray}\label{eq:C9mu}
C^{\mu,{\rm NP}}_{9} &=  &- \frac{\pi}{\alpha_{em} V_{tb}V_{ts}^* } \lambda_{33}^q \lambda_{\mu\mu}^\ell [V_{ts}^*\theta_q e^{-i \phi_q}+r_{23}]\\
&&\qquad\qquad  \times\left(C_T +C_S +C_{LR}' \right)
\,,\nonumber\\
C^{\mu,{\rm NP}}_{10}&=  &- \frac{\pi}{\alpha_{em} V_{tb}V_{ts}^* } \lambda_{33}^q \lambda_{\mu\mu}^\ell [V_{ts}^*\theta_q e^{-i \phi_q}+r_{23}] \\
&&\qquad\qquad  \times \left(- C_T -C_S +C_{LR}' \right)  \,, \nonumber\\
C^{\mu,{\rm NP}}_{9'} &=  &- \frac{\pi}{\alpha_{em} V_{tb} V_{ts}^*} \lambda_{33}^q \lambda_{\mu\mu}^\ell r_{23} \left( C_{RR}'+C_{RL}'   \right) \,,  \\
C^{\mu,{\rm NP}}_{10'} &=  &- \frac{\pi}{\alpha_{em} V_{tb} V_{ts}^*} \lambda_{33}^q \lambda_{\mu\mu}^\ell r_{23}  \left(C_{RR}'  - C_{RL}' \right) \, ,
\label{eq:C10pmu}
\end{eqnarray}
and in the SM, $C_{9}^{\ell,{\rm SM}}=4.07$ and $C_{10}^{\ell,{\rm SM}}=-4.31$ at $\mu=m_b$ for all three lepton flavours, whereas chirally flipped operators are negligible.

The rare B decays $B \to K^{(*)} \nu\bar\nu$ can be conveniently expressed in presence of NP of the form in Eq.~\eqref{eq:ops} as~\cite{Altmannshofer:2009ma}
\begin{align}
\mathcal B(B \to K \nu \bar \nu ) =&  (4.5\pm0.7) \times 10^{-6} \frac{1}{3}\sum_\nu (1 - 2 \eta_\nu) \epsilon_\nu^2\,,  \nonumber \\
\mathcal B(B \to K^* \nu \bar \nu ) =&  (6.8\pm1.1) \times 10^{-6} \frac{1}{3}\sum_\nu (1 + 1.31 \eta_\nu) \epsilon_\nu^2\,,  \nonumber \\
\mathcal B(B \to X_s \nu \bar \nu ) =&  (2.7\pm0.2) \times 10^{-5} \frac{1}{3}\sum_\nu(1 + 0.09 \eta_\nu) \epsilon_\nu^2\,,  \nonumber \\
\langle F_L \rangle =&  (0.54\pm 0.01) \frac{\sum_\nu (1 + 2 \eta_\nu)\epsilon_\nu^2}{\sum_\nu (1+1.31 \eta_\nu)\epsilon_\nu^2}\,, 
\label{eq:BKnunu}
\end{align}
where $\langle F_L \rangle$ is the longitudinal $K^*$ polarisation fraction in $B\to K^* \nu \bar \nu$ decays.   
For each flavour of neutrino $\nu=\nu_e,\nu_\mu,\nu_\tau$, the two NP parameters can in turn be expressed as
\beq
\epsilon_\nu = \frac{\sqrt{|C^\nu_L|^2 + |C^\nu_R|^2}}{|C^\nu_{\rm SM}|}\,, ~ \eta_\nu = \frac{- {\rm Re}(C^\nu_L C^{\nu*}_R)}{|C^\nu_L|^2 + |C^\nu_R|^2}\,,
\eeq
where $C_{L,R}^{\nu} = C^{\nu,{\rm SM}}_{L,R} + C^{\nu,{\rm NP}}_{L,R} $ and
$C^{\nu,{\rm SM}}_L = -6.38$ and $C^{\nu,{\rm SM}}_R=0$ at $\mu=m_b$. Including leading $U(2)_q$ breaking effects we can write again 
\begin{align}
C^{\nu_\alpha, {\rm NP}}_L &=   -\frac{\pi}{\alpha_{em} V_{tb}V_{ts}^*}\lambda_{33}^q \lambda_{\alpha\alpha}^\ell [V_{ts}^*\theta_q e^{-i\phi_q} +r_{23}] [C_S-C_T]\,, \label{eq:CnuL} \\
 C^{\nu_\alpha, {\rm NP}}_R &= -\frac{ \pi}{\alpha_{em} V_{tb} V_{ts}^*}\lambda_{33 }^q \lambda_{\alpha\alpha}^\ell r_{23} C'_{RL}\,,
 \end{align}
 with $\alpha=e,\mu,\tau$.
We wrote these expressions neglecting the small neutrino mass effects setting effectively $U^{\rm PMNS}$ to the identity matrix. Note that any deviations from SM in $\langle F_L \rangle$ or non-universal deviations in $\mathcal B(B \to (K,K^*,X_s) \nu \bar \nu ) / \mathcal B(B \to (K,K^*,X_s) \nu \bar \nu )_{\rm SM}$ would signal the presence of right-handed quark currents ($C'_{RL}\neq0$) and thus departures from the  (G)MFV limit. 

Similarly, the rare kaon decays $K^+ \to \pi^+ \nu\bar\nu$ and $K_L \to \pi^0 \nu\bar\nu$ can be conveniently expressed in presence of NP of the form in Eq.~\eqref{eq:ops} as~\cite{Kamenik:2017tnu}
\begin{align}
& \mathcal B(K^+ \to \pi^+ \nu \bar \nu (\gamma)) =  (8.4\pm1.0) \times 10^{-11}  \nonumber \\
 & \times \frac{1}{3} \sum_\nu \left| 1+ \frac{ C^{\nu,\rm NP}_{sd}}{V_{ts} V_{td}^* X_{t} + (X_c + \delta X_{c,u})  V_{cs} V_{cd}^*} \right|^2\,,\nonumber \\
& \mathcal B(K_L \to \pi^0 \nu \bar \nu ) =  (3.4\pm0.3) \times 10^{-11}  \nonumber \\
 & \times \frac{1}{3} \sum_\nu \left[1+{\rm Im} \left(\frac{ C^{\nu,\rm NP}_{sd}}{V_{ts} V_{td}^* X_{t}} \right)\right]^2\,,
 \label{eq:Kpinunu}
\end{align}
where $X_i$ are defined in Ref.~\cite{Brod:2010hi} and $s_W \equiv \sin \theta_W \simeq 0.48$, $c_W \equiv \cos \theta_W$. Numerically, $X_t = 1.469(17) $~\cite{Brod:2010hi} and  $(X_c + \delta X_{c,u})  = 0.00106(6) $~\cite{Isidori:2005xm, Brod:2008ss}\,.  For each neutrino flavour $\nu=\nu_e,\nu_\mu,\nu_\tau$,
$C^{\nu,\rm NP}_{sd}$ receives contributions from three operators of the weak effective Hamiltonian yielding
\begin{align}
& C^{\nu_\alpha,{\rm NP}}_{sd} =   \frac{\pi s_W^2}{\alpha_{em}}\lambda_{33}^q  \lambda_{\alpha\alpha}^\ell  [\theta_q^2 V_{ts} V_{td}^* \left( C_S -C_T\right) \nonumber\\
 & + \theta_q ( V_{ts} e^{i \phi_q} r^*_{13} + V^*_{td} e^{-i \phi_q} r_{23}) \left( C_S -C_T\right) \nonumber\\
 &  +r_{12} \left( C_S -C_T + C'_{RL}\right)  ]\,, \label{eq:Cnu}
\end{align}
 where $\alpha=e,\mu,\tau$, we have again neglected neutrino mass effects. Since $s\to d $ transitions only appear at quadratic order in GMFV breaking of $U(2)_q$, in this case we have included up to quadratic $U(2)_q$ breaking terms, but at the same time kept only the linear $U(2)_q$ breaking contributions {\it beyond} GMFV, since these suffice for our following discussion. 

A short comment regarding the recent intriguing results on the $K_L \to \pi^0 \nu \bar \nu$ from the KOTO collaboration is in order at this point. As noted in Ref.~\cite{Kitahara:2019lws}, at the 68\% CL, the result, if combined with the NA62 bound on $K^+ \to \pi^+ \nu\bar\nu$, violates the Grossman-Nir bound~\cite{Grossman:1997sk} and cannot be explained without invoking isospin breaking NP~\cite{He:2020jzn, 1794494} and additional long-lived neutral final states in the $K_L$ decay beyond the three SM neutrinos, see e.g. Ref~\cite{Jho:2020jsa,Liu:2020qgx,Ziegler:2020ize}. 
None of the NP scenarios we consider can thus fully accommodate both measurements. At most we can comment on ways to approach the Grossman-Nir bound. In particular, we note that new CP phases  in  $s\to d$ transitions only appear beyond the (G)MFV limit~\cite{Kagan:2009bn}. In the case of $K\to \pi \nu\bar \nu$ decays we can see this explicitly in Eq.~\eqref{eq:Cnu} since only terms proportional to $r_{ij}$ may carry additional phases. These terms should thus dominate over the first row indicating large departures from the (G)MFV limit. Unfortunately, little can be said about the implications of $b\to s\mu\mu$ data model independently in this part of parameter space. A potential future experimental confirmation of $C_{9\prime}^{\mu,NP} \neq 0$  
could at best provide circumstantial evidence for the presence of $U(2)_q$ breaking beyond (G)MFV. 
 
\begin{figure}[t!]
\centering
\includegraphics[width=7.cm]{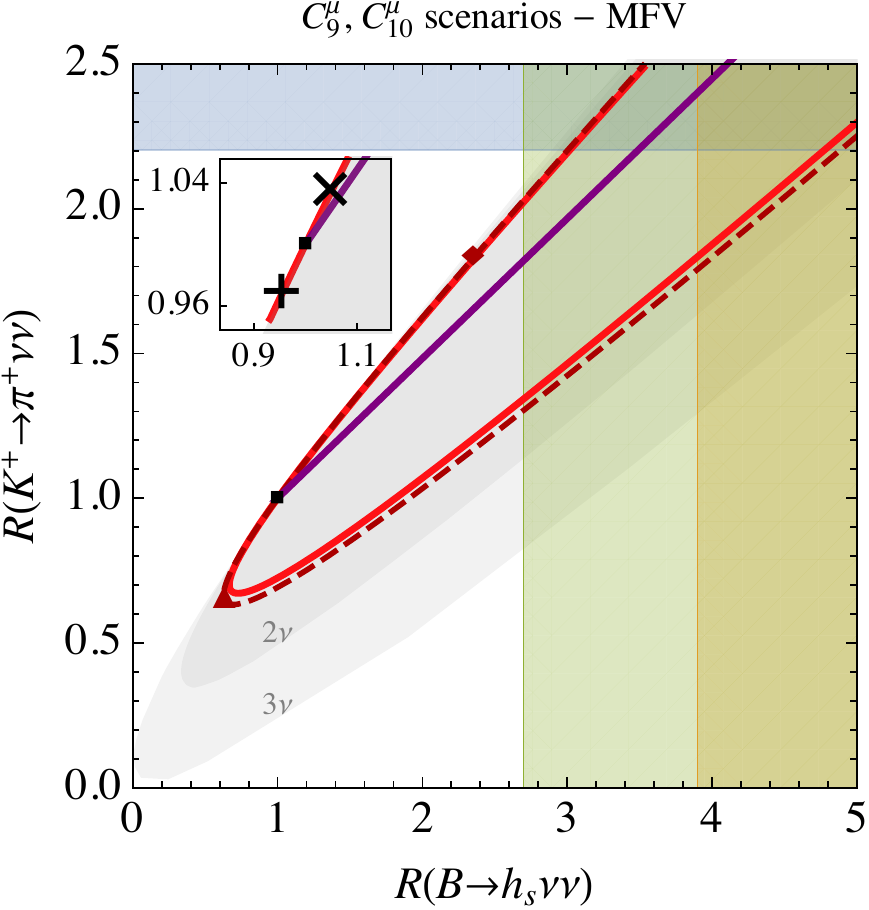}
\caption{Correlation between the ratios $R(B\to h_s\nu\bar\nu)$ ($h_s=K,K^*,X_s$) and $R(K^+\to \pi^+\nu\bar\nu)$ in the linear MFV limit.  The SM value is represented by the black square. The region allowed for arbitrary NP effects in $\nu_\mu$ and $\nu_\tau$ only (all three neutrino flavours) is show in dark grey (light grey respectively). Curves are drawn for the specific $U(1)^3_\ell$ scenarios 1 (NP only in muons, red), 2 (opposite NP effects in muons and taus, purple) and 3 (hierarchical NP effects according to the generation, dashed brown).
Scenarios with $C_S=0$ or $C_T=0$ are indicated as black $\times$ and $+$ respectively in the inset plot for scenario 1 and brown $\Diamond$ or $\bigtriangleup$ respectively  for scenario 3. { The horizontal and vertical bands correspond to the 90 \% CL limits on the observables for $R(B\to K^*\nu\bar\nu)$ (green), $R(B\to K\nu\bar\nu)$  (orange)~\cite{Grygier:2017tzo} and $R(K^+\to \pi^+\nu\bar\nu)$~\cite{Ruggiero} (blue).}
}
\label{fig:C9C10}
\end{figure}

\mysection{Results in the linear MFV case} We first consider the limit of (linear) MFV in which $b\to s\nu\bar\nu$ and $s\to d\nu\bar\nu$ FCNC transitions are rigidly correlated via the corresponding CKM prefactors in Eqs.~\eqref{eq:CnuL} and~\eqref{eq:Cnu} and $C'_{RL}=C'_{RR}=0$. Even before considering the implications of $R_{K^{(*)}}$,  this immediately implies a very general correlation between $B \to h_s \nu\bar\nu$ 
and $K\to\pi \nu \bar\nu$ rates, driven by the combination of Wilson coefficients $C_S - C_T$ in Eq~\eqref{eq:ops}. For conciseness, we consider the branching ratios normalised to their SM values by introducing $R(i\to f) \equiv \mathcal B(i\to f) /  \mathcal B(i\to f)_{\rm SM}$. The allowed region for these ratios is shown shaded in  darker $(2\nu)$ and  lighter $(3\nu)$ grey in Fig.~\ref{fig:C9C10}, where arbitrary MFV NP effects in  two $(2\nu)$ or three $(3\nu)$ neutrino flavours (with arbitrary $\lambda^q_{33} \lambda^{\ell}$) have been considered, respectively. The two ratios $R$ are bounded by the same minimal value $(1-N_\nu/3)$ where $N_\nu$ is the number of neutrino flavours affected by NP. 
Also shown are the present experimental constraints coming from NA62~\cite{Ruggiero} and $B$-factories~\cite{Grygier:2017tzo} respectively.  An interesting observation is that a pair of future $B \to h_s \nu\bar\nu$ and $K\to\pi \nu \bar\nu$ rate measurements outside of this (albeit large) region would be a clear indication of non-MFV NP.  On the same plot we also superimpose the three specific $U(1)^3_\ell$ scenarios.

In the $b \to s \ell^+ \ell^-$ analysis, the MFV limit corresponds to the $(C_9^{\mu,\rm NP}$, $C_{10}^{\mu,\rm NP})$ scenario. In terms of the EFT operator basis in Eq~\eqref{eq:ops}  $R_{K^{(*)}}$ measurements (and more generally $b\to s\ell\ell$ data) favour non-zero values for both $C_S+C_T$ and $C'_{LR}$. Since $B \to h_s \nu\bar\nu$ and $K\to\pi \nu \bar\nu$ depend on the orthogonal $C_S-C_T$ combination, interesting implications can only be derived in specific scenarios allowing us to convert the information from $b \to s \ell^+ \ell^-$ observables into a constraint on $C_S$ and $C_T$. The simplest possibilities ($C_S=0$ or $C_T=0$) are indicated in Fig.~\ref{fig:C9C10} for $U(1)^3_{\ell}$ scenarios 1 and 3 respectively. On the other hand, in scenario 2, no significant deviations are expected in either case.  We observe that the pure $SU(2)$ triplet ($C_S=0$) scenario 3 ($\lambda^{\ell}_{\tau\tau} / \lambda^{\ell}_{\mu\mu} = m_\tau / m_\mu$) was already close to being probed by searches for $B \to K^{(*)} \nu\bar\nu$ at the $B$-factories. The final projected sensitivity of Belle II could be sufficient to eventually also distinguish between the pure $SU(2)$ triplet ($C_S=0$) and singlet ($C_T=0$) limits of scenario 1.

\begin{figure}[t!]
\centering
\includegraphics[width=7.cm]{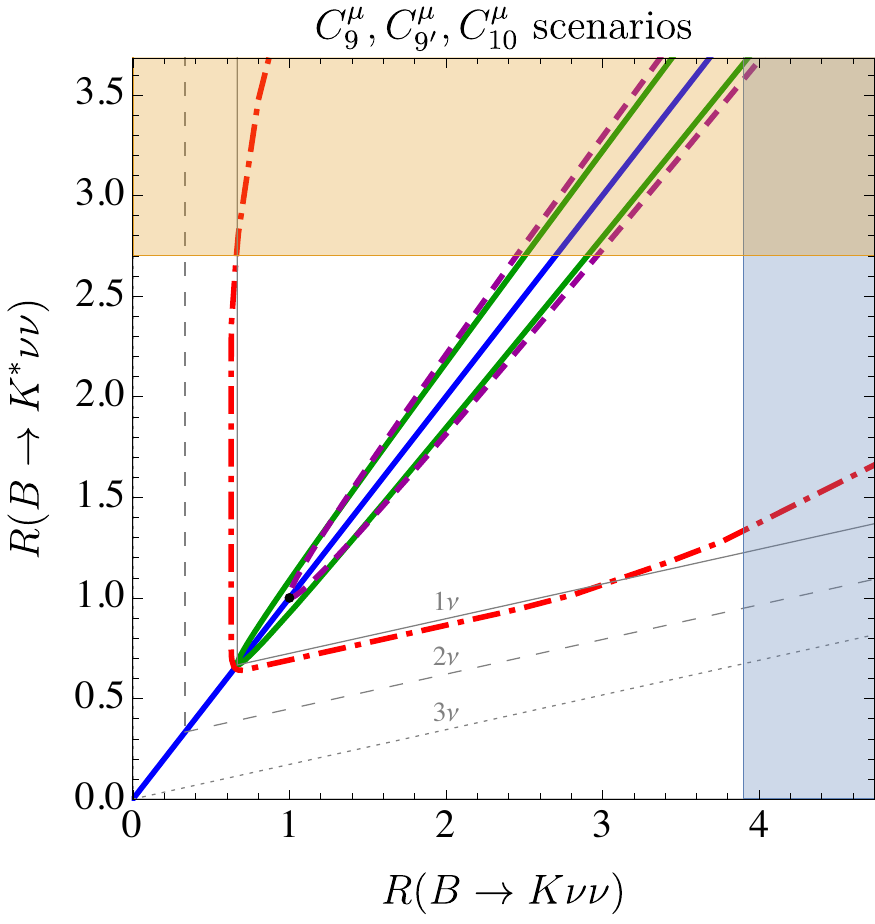}
\caption{
Correlation between the  ratios  $R(B\to K\nu\bar\nu)$ and $R(B\to K^*\nu\bar\nu)$ in the presence of NP
in  $b\to s\mu\mu$ transitions through $(C^{\mu,\rm NP}_9, C^{\mu,\rm NP}_{9'})$. 
The diagonal blue line corresponds to the (G)MFV case. The 1 $\sigma$ region allowed by $b\to s\mu\mu$ transitions 
yields an allowed region depending on the assumption on the couplings to leptons, inside the solid green line
for scenario 1 (NP only in muons), dashed purple for scenario 2 (opposite NP effects in muons and taus) and dot-dashed red for scenario 3 (hierarchical NP effects according to the generation).
Without information on the size of the right-handed FCNCs from $b\to s \mu^+ \mu^-$, the allowed region assuming significant NP couplings to 1, 2, 3 neutrinos is above and on the right of the solid, dashed, dotted grey contours, respectively. The horizontal and vertical bands correspond to the 90 \% CL limits on the observables for $R(B\to K^*\nu\bar\nu)$ (orange) and $R(B\to K\nu\bar\nu)$ (blue)~\cite{Grygier:2017tzo}.}
\label{fig:U2}
\end{figure}

\mysection{Results with right-handed currents}
Beyond the linear MFV limit any correlation between $b\to s$ and $s\to d$ FCNCs is lost in general. Nonetheless, the potential presence of right-handed $b\to s$ FCNCs in the $(C^{\mu,\rm NP}_9, C^{\mu,\rm NP}_{9'})$ scenario as well as the leptonic flavour structure of NP can both still be probed using correlations among two $B \to h_s \nu\bar\nu$  modes, as shown in Fig.~\ref{fig:U2} for the case $R(B \to K^{} \nu\bar\nu)$ vs. $R(B \to K^{*} \nu\bar\nu)$.  

First note that in the MFV limit relative NP effects in both modes are expected to be identical as indicated by the  diagonal red line. Beyond MFV however, the amount of deviation from the diagonal would directly indicate the number of lepton flavours affected by NP. 

In scenario 1 (where only muons couple significantly to NP)  and scenario 2 (where muons and taus have opposite NP couplings) the $b\to s\ell\ell$ fit 
for $(C^{\mu,\rm NP}_9, C^{\mu,\rm NP}_{9'})$
singles out a narrow region around the diagonal in this plane, whereas scenario 3 leaves a much larger region allowed.
Conversely, a measurement of the two $b\to s \nu\bar \nu$ modes outside of the region for scenario 1 would indicate significant (right-handed FCNC) NP couplings to other neutrino species, e.g. $\nu_\tau$. 

In absence of information on the size of the right-handed FCNCs from the $b\to s \mu^+ \mu^-$ modes in principle the whole region within the grey $1\nu$ contour could be accessible, with limits corresponding $\eta_\nu=-1/2$ and $+1/2$ (MFV corresponding to $\eta_\nu=0$). In presence of significant couplings also to tau neutrinos as e.g. in scenario 3, the whole region within the grey dashed $2\nu$ contour is possible, even when the existing constraints coming from $b\to s \mu^+ \mu^-$ modes are taken into account. 

Finally, in presence of significant right-handed FCNCs coupling to all three neutrino  flavours the whole region within the grey dotted $3\nu$ contour would be possible in principle.

\mysection{Conclusions}
In this article, we have investigated the consequences of deviations from the SM observed in $b\to s\mu\mu$ transitions for FCNC processes involving down-type quarks and neutrinos. Motivated by the results from the global fits to $b\to s\ell\ell$ observables as well as measurements and bounds on FCNC processes with neutrinos, we have considered a general EFT description of FCNC transitions in terms of $SU(2)_L$ gauge invariant operators including those with right-handed quarks and charged leptons. This allowed us to describe with the same short-distance Wilson coefficients $b\to s\mu\mu$, $b\to s\nu\bar\nu$ and $s\to d\nu\bar\nu$. 

We have briefly touched upon the status of $K_L\to\pi^0\nu\bar\nu$, which is only affected by CPV NP, requiring new flavour dynamics beyond (G)MFV. In this case, there is no clear correlation with the other FCNC modes discussed here. The recent KOTO results that violate the Grossman-Nir bound are particularly challenging to explain in conjunction with the NA62 bound on $K^+\to \pi^+ \nu\bar\nu$, and they cannot be accommodated within our framework.

Assuming (G)MFV in the quark sector, we have studied the correlation between the branching ratios for $B\to h_s\nu\bar\nu$ and $K^+\to\pi^+\nu \bar\nu$. Such a correlation is already present without assuming any specific structure for the neutrino NP couplings, but it can be made even more precise once specific NP scenarios assign  specific values to these couplings. Moreover, for scenarios with no triplet ($C_T=0$) or singlet ($C_S=0$) contributions, the fits to $(C_9^{\mu,{\rm NP}},C_{10}^{\mu,{\rm NP}})$ can be immediately converted into predictions for these two branching ratios in terms of ${\bf R}_{\nu\nu} \equiv [R(B\to h_s\nu\bar\nu),R(K^+\to\pi^+\nu\bar\nu)]$. { In scenario 1 where NP couples only to muons, we find ${\bf R}_{\nu\nu} \simeq (0.95, 0.97)$ if $C_S=0$ and ${\bf R}_{\nu\nu} \simeq (1.05,1.03)$ if $C_T=0$. In scenario 2 where muons and taus have opposite couplings, the values remain very close to the SM. In scenario 3 where NP hierarchical couplings proportional to the lepton mass are assumed, we find ${\bf R}_{\nu\nu} \simeq (0.64,0.65)$ if $C_S=0$ and ${\bf R}_{\nu\nu} \simeq (2.4,1.8)$ if $C_T=0$.}

Moving beyond the (G)MFV limit, we have investigated the correlation between $B\to K\nu\bar\nu$ and $B\to K^*\nu\bar\nu$, in particular showing that depending on the NP lepton couplings
 also the scenario with NP in $(C_9^{\mu,{\rm NP}},C_{9'}^{\mu,{\rm NP}})$ can yield a tight correlation between the two modes when the  $b\to s\ell\ell$ measurements are taken into account. For example, in scenarios 1 and 2, the ratio $R(B\to K\nu\bar\nu) / R(B\to K^*\nu\bar\nu)$ cannot deviate from unity by more than 8\%.  More generally however, such measurements could establish NP flavour breaking beyond (G)MFV as well indicate the number of lepton flavours affected by NP. 

We hope that our results will strengthen the case for more accurate measurements of $b\to s\nu\bar\nu$ and $s\to d\nu\bar\nu$ modes, in order to determine which direction should be followed to develop viable NP models describing the hints of deviations in $b\to s\mu\mu$ and providing a viable connection with other quark generations at the same time.
 
\begin{acknowledgments}

SF and JFK acknowledge the financial support from the Slovenian Research Agency (research core funding No. P1-0035 and J1-8137).

\end{acknowledgments}

\bibliographystyle{elsarticle-num}

\bibliography{current}

\end{document}